\documentclass[preprint,12pt]{elsarticle}

\usepackage{amssymb}
\usepackage{amsmath}
\usepackage{tabularx}  % Add to preamble
\usepackage{array}     % For m{} column type
\usepackage{ragged2e}  % For better text justification
\usepackage{eurosym}
\usepackage{multirow}
\usepackage{makecell}
\usepackage{xurl}

\journal{ }

\begin{document}

\begin{frontmatter}



\title{Machine-learning competition to grade EEG background patterns in newborns with hypoxic-ischaemic encephalopathy}


\author[label1,label2,label3]{Fabio Magarelli}
\author[label1,label2]{Geraldine B. Boylan}
\author[label4]{Saeed Montazeri}
\author[label5]{Feargal O'Sullivan}
\author[label5]{Dominic Lightbody}
\author[label6]{Minoo Ashoori}
\author[label7]{Tamara Skoric}
\author[label1,label2,label8]{John M. O'Toole}

\affiliation[label1]{
    organization={INFANT Research Centre, University College Cork},
    city={Cork},
    country={Ireland}
}
\affiliation[label2]{
    organization={Department of Pediatrics and Child Health, University College Cork},
    city={Cork},
    country={Ireland}
}
\affiliation[label3]{
    organization={The SFI Centre for Research Training in Artificial Intelligence, CRT in AI},
    country={Ireland}
}
\affiliation[label4]{
    organization={Department of Physiology, University of Helsinki},
    city={Helsinki},
    country={Finland}
}
\affiliation[label5]{
    organization={Embedded Systems, University College Cork},
    country={Ireland}
}
\affiliation[label6]{
    organization={Department of Physiology, University College Cork},
    country={Ireland}
}
\affiliation[label7]{
    organization={Faculty of Technical Sciences, University of Novi Sad},
    country={Serbia}
}
\affiliation[label8]{
    organization={CergenX Ltd},
    country={Ireland}
}

%% Abstract
\begin{abstract}
Machine learning (ML) has the potential to support and improve expert performance in monitoring the brain function of at-risk newborns. Developing accurate and reliable ML models depends on access to high-quality, annotated data, a resource in short supply. ML competitions address this need by providing researchers access to expertly annotated datasets, fostering shared learning through direct model comparisons, and leveraging the benefits of crowdsourcing diverse expertise. We compiled a retrospective dataset containing 353 hours of EEG from 102 individual newborns from a multi-centre study. The data was fully anonymised and divided into training, testing, and held-out validation datasets. EEGs were graded for the severity of abnormal background patterns. Next, we created a web-based competition platform and hosted a machine learning competition to develop ML models for classifying the severity of EEG background patterns in newborns. After the competition closed, the top 4 performing models were evaluated offline on a separate held-out validation dataset. Although a feature-based model ranked first on the testing dataset, deep learning models generalised better on the validation sets. All methods had a significant decline in validation performance compared to the testing performance. This highlights the challenges for model generalisation on unseen data, emphasising the need for held-out validation datasets in ML studies with neonatal EEG. The study underscores the importance of training ML models on large and diverse datasets to ensure robust generalisation. The competition’s outcome demonstrates the potential for open-access data and collaborative ML development to foster a collaborative research environment and expedite the development of clinical decision-support tools for neonatal neuromonitoring.

\end{abstract}


\begin{highlights}
\item Machine learning competition to grade newborn EEG background abnormalities
\item Open-source platform to host machine learning competitions
\item Crowdsourced machine learning and open data to advance newborn EEG analysis
\item Diverse, high-quality datasets essential for robust machine learning models
\end{highlights}

%% Keywords

\end{frontmatter}



\section{Introduction}
\label{sec1}
The severity of background patterns in EEG recordings obtained over the first days after birth reflects the current level of neurological injury and is indicative of longer-term neurodevelopmental outcomes \cite{Moghadam2022-aw} \cite{Murray2016} \cite{Menache2002}. It offers a valuable cot-side evaluation of brain function and can assist in neurocritical care. Not all neonatal care centres, however, have access to the necessary expertise for interpreting EEG recordings. Machine learning (ML) tools to automatically analyse EEG data could assist expert review by reducing subjective interpretation and greatly improving efficiency and effectiveness, allowing for EEG review on a larger scale \cite{Ryan2024}.

Numerous machine learning models \cite{Moghadam2022-aw} \cite{Raurale2021} \cite{Roy2019} \cite{Ansari2016} \cite{Matic2014} have been developed to classify EEG severity. Notably, these models are developed by researchers in research centres, often with limited training datasets, and are seldom deployed for clinical validation. ML competitions serve as a fair benchmarking tool for comparing models using the same training, testing, and validation data. Competitions have the potential to cultivate collaborative partnerships among researchers, promote knowledge exchange, and enable competition hosts to leverage collective expertise to enhance outcomes and drive advancements within the field  \cite{Banachewicz2022-br} \cite{9202557}.

In 2024, the machine learning (ML) competition landscape experienced significant growth, hosting over 400 competitions across more than 20 platforms, with total prize money exceeding \$22 million. Kaggle maintained its leadership position, surpassing 22 million registered users and offering over \$4 million in prizes while CodaLab launched the highest number of competitions (113 out of 400). Additionally, Hugging Face launched an updated version of its competition platform in 2024, with new features. Other notable platforms include DrivenData, EvalAI \cite{https://doi.org/10.48550/arxiv.1902.03570}, with an ever growing list of new sites \cite{carlens2024state} \cite{carlens2025state}. These developments underscore the dynamic and evolving nature of the ML competition ecosystem, highlighting both opportunities and challenges for participants and organizers alike. Yet these commercial platforms can incur high costs for those hosting the competition and may provide little control of the data.

We chose to develop our own platform for hosting ML competitions. This decision gave us more flexibility and control over the development process and enabled us to develop an open-access platform without fees. This platform is open for anyone to use to create their own ML competition. We deployed the platform on a university server to host a competition to grade the severity of neonatal EEG background patterns in newborns with hypoxic-ischaemic encephalopathy (HIE).

In the following sections, we present the details of our ML competition, including the dataset used, the bespoke platform for hosting the competition, and the evaluation of submissions. We discuss the top-performing models, their generalizability on unseen data, and the implications of our findings for future research and clinical applications in neonatal care

\section{Methods}
\label{sec2}
We developed and ran a ML competition to solicit automated methods to classify 4 grades of background EEG patterns in newborns with HIE. To do so, we created a bespoke web platform to host machine learning competitions. The competition had the following objectives:

\begin{itemize}
    \item Evaluate performances of different ML methods and identify common characteristics for best performance;
    \item Identify suitable candidate ML models for development of prototype models for further clinical validation;
    \item Promote awareness of open-access EEG data with annotations;
    \item Promote awareness of open challenges in ML research for automated analysis of neonatal EEG.
\end{itemize}

Our goal is to encourage more ML experts to apply their skills to neonatal EEG, an area with real potential for positive clinical impact. Although this is an active research field with a dedicated but small research community, expanding this community to include those with expertise in applying ML solutions to other domains has the potential to build critical mass and achieve the significant breakthroughs required for the translation of ML methods to enhance newborns’ neurocritical care.

\subsection{EEG Data}
\label{subsec2.1}
The competition used retrospective EEG data recorded as part of the multi-centre study (ANSeR), including a clinical trial to investigate the utility of an ML seizure detection model \cite{Pavel2020-zd} \cite{Rennie2019-vn}. Continuous EEG (cEEG) was obtained from newborns with gestational age (GA)  $\ge$ 36 weeks who were at high risk of seizure and required cEEG monitoring. Ethical approval was obtained from the Cork Ethics Research Committee for secondary use of existing anonymised data. The same Committee also approved the open access publication of the anonymised EEG recordings that were used in the competition. Permission to share this data was also obtained from the Data Protection Officer at University College Cork, Ireland. The competition participants only had access to the published open-access portion of the data \cite{OToole2023} \cite{o_toole_2022_6587973}.

The  EEG data was recorded from newborns with a confirmed HIE diagnosis in the ANSeR studies \cite{Pavel2020-zd} \cite{Rennie2019-vn} \cite{https://doi.org/10.1111/epi.17468}. The EEG data was collected within the first 48 hours of birth using NicoletOne ICU Monitor (Natus, USA), Nihon Kohden EEG (Neurofax EEG-1200, Japan) or XLTek EEG (Natus, USA) with a reduced electrode 10-20 system using the channels: F3, F4, C3, C4, Cz, T3, T4, O1/P3 and O2/P4. The recordings had a sampling rate of 250Hz or 256Hz. A 1-hour epoch was extracted from the continuous recording at specific time points (6 hr, 12Hr, 24Hr, 36Hr and 48Hr from birth, where available) \cite{Raurale2021} \cite{OToole2023} \cite{o_toole_2022_6587973} \cite{https://doi.org/10.1111/epi.17468} \cite{OSULLIVAN2023118917}. The grading scheme used to classify the background pattern of patterns was that proposed by Murray and colleagues, where grades 0 and 1 were merged into a single grade  (grade 1 in Table \ref{tab:eeg_grades}) \cite{Murray2016} \cite{Murray2009}:

\begin{table}[t]
\centering
\begin{tabular}{p{0.1\textwidth} p{0.3\textwidth} p{0.5\textwidth}}
\hline
\textbf{Grade} & \textbf{EEG Findings} & \textbf{Description} \\
\hline
1 & Normal or Mildly abnormal & Continuous activity with normal or mildly abnormal activity (e.g., mild asymmetry, mild voltage depression or poorly defined SWC) \\
\hline
2 & Moderately abnormal & Discontinuous activity with IBI $>$ 10s, no clear SWC or clear asymmetry or asynchrony \\
\hline
3 & Major abnormalities & IBI of 10-60s, severe background activity attenuation or no SWC \\
\hline
4 & Inactive EEG & Background activity $<$ 10µV or severe discontinuity with IBI $>$ 60s \\
\hline
\end{tabular}
\caption{Classification of EEG Background activity according to Murray and colleagues' scheme, 2016 \cite{Murray2016} where grade 0 and 1 have been merged to facilitate ML models training. SWC: Sleep Wake Cycle; IBI: Inter Burst Interval.}
\label{tab:eeg_grades}
\end{table}

\begin{figure}[t]
\centering
\includegraphics[width=\textwidth]{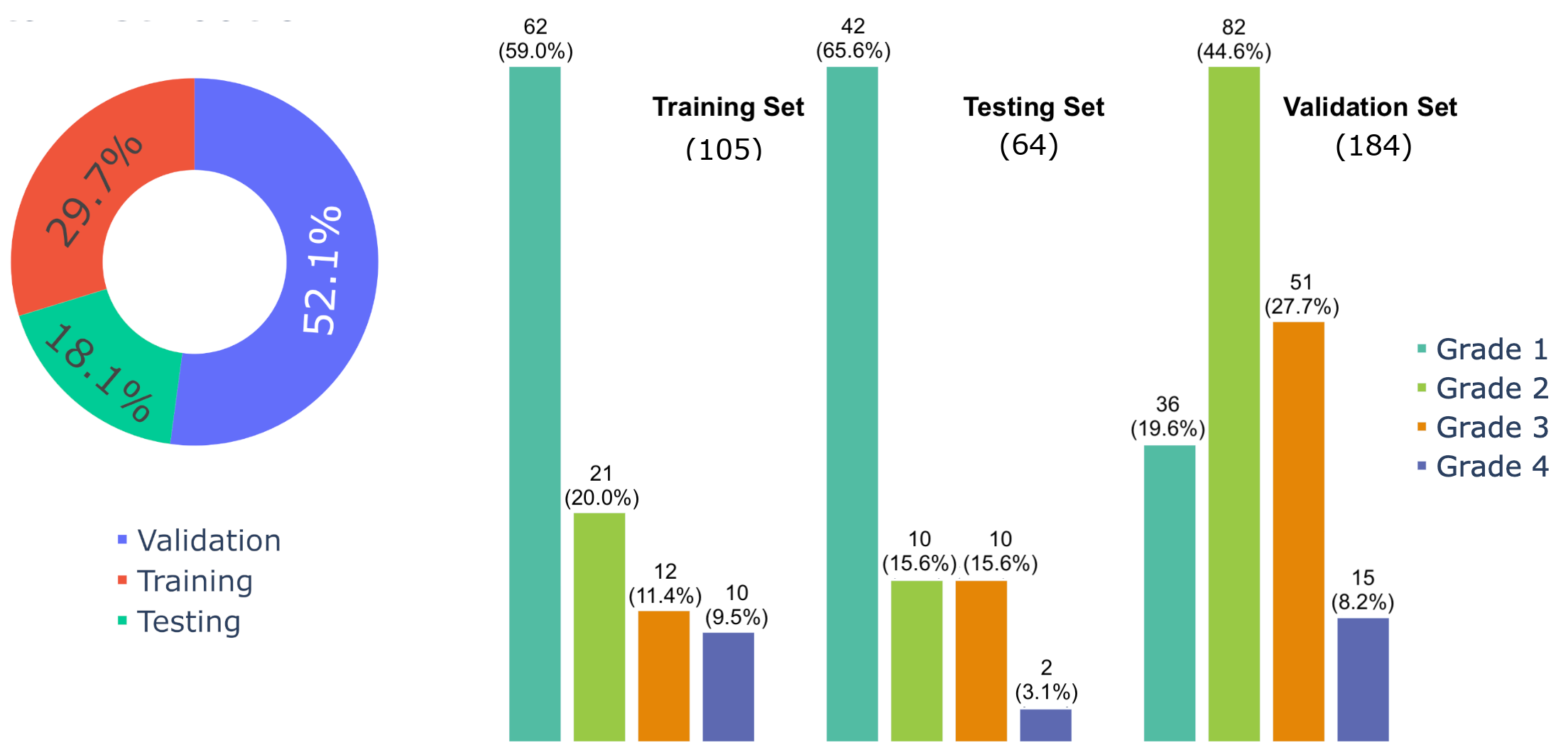}
\caption{(Left) Data distribution within the different datasets; (Right) distribution of EEG grades.}
\label{fig:data-distrib}
\end{figure}

The data was then split into 3 datasets (Figure \ref{fig:data-distrib}):

1. \textbf{Public dataset:} It comprises 105 (62\%) labelled epochs for training individual submissions (from now on referred to as training data or dataset) and 64 (38\%) unlabelled epochs for the platform’s automatic testing (from now on referred to as testing data or dataset).
2. \textbf{Validation dataset:} (hidden from the participants and unpublished) – used for evaluation after the competition was closed

The public dataset consists of a fully anonymised, open-access dataset \cite{o_toole_2022_6587973}. At the time of competition, only the labels (EEG grades) for the training part of this dataset were publicly available. During the competition, the participants developed their methods on the training data, which consisted of the labelled portion of the public dataset. They then could test their models using the unlabelled portion of the public dataset. Feedback on performance was provided by submitting the predicted grades to the competition platform, and results were presented on a leaderboard. We split the data on newborns rather than epochs to avoid data leakage between datasets.

The EEG in the public dataset was recorded at Cork University Maternity Hospital (CUMH). This dataset was randomly divided into approximately 60:40 for training–testing sets. The training portion contained 105 epochs from 30 newborns and was used by the participants to train and evaluate their models. The testing dataset comprised of 64 epochs from 23 newborns. The validation dataset included 184 1-hour epochs from 49 newborns. This dataset was not available to the competition participants. This dataset was chosen to evaluate generalisation performance after the ML models were frozen and was selected to be closer to a real-world scenario: EEGs from an international multi-centre dataset were included, of which only a subset of newborns were from Cork University Maternity Hospital \cite{o_toole_2022_6587973} \cite{Pavel2020-zd}. Demographics and clinical characteristics for the 2 datasets are present in Table \ref{tab:clinical_chars}.

\begin{table}[t]
\centering
\makebox[\textwidth]{
\begin{tabular}{>{\raggedright}p{0.4\textwidth} 
                >{\centering}p{0.25\textwidth} 
                >{\centering}p{0.25\textwidth} 
                >{\centering\arraybackslash}p{0.15\textwidth}}
\hline
 & \textbf{Public Dataset} & \textbf{Validation Dataset} & \textbf{p-value} \\
\hline
\textbf{Number of newborns} & 53 & 49 & \\
\textbf{Gestational age (weeks)} & \begin{tabular}{@{}c@{}}40.0\\(39.4 to 40.7)\end{tabular} & \begin{tabular}{@{}c@{}}40.4\\(39.4 to 41.4)\end{tabular} & 0.153 \\
\textbf{Birth weight (g)} & \begin{tabular}{@{}c@{}}3,470\\(3,190 to 3,800)\end{tabular} & \begin{tabular}{@{}c@{}}3,379\\(3,110 to 3,820)\end{tabular} & 0.574 \\
\textbf{Lowest cord pH} & \begin{tabular}{@{}c@{}}7.03\\(6.88 to 7.19) \textdagger\end{tabular} & \begin{tabular}{@{}c@{}}7.0\\(6.90 to 7.15) \textdagger\end{tabular} & 0.895 \\
\textbf{Apgar score (10min)} & \begin{tabular}{@{}c@{}}5 (4 to 8) \textdaggerdbl\end{tabular} & \begin{tabular}{@{}c@{}}5 (4 to 7.5) \textdaggerdbl\end{tabular} & 0.277 \\
\textbf{Sex (male)} & 33 (62\%) & 29 (59\%) & 0.840 \\
\hline
\multicolumn{4}{p{\dimexpr\textwidth-4\tabcolsep}}{\textbf{Final Diagnosis (HIE):}} \\
\emph{Mild} & 24 (45\%) & 16 (33\%) & 0.229 \\
\emph{Moderate} & 20 (38\%) & 26 (53\%) & 0.774 \\
\emph{Severe} & 9 (17\%) & 7 (14\%) & 0.322 \\
\hline
\multicolumn{4}{p{\dimexpr\textwidth-4\tabcolsep}}{\textbf{Therapeutic hypothermia}} \\
\emph{Cooled} & 31 (58\%) & 46 (94\%) & $<$0.001 \\
\hline
\multicolumn{4}{p{\dimexpr\textwidth-4\tabcolsep}}{\textbf{Most intensive resuscitation required:}} \\
\emph{Facial Oxygen} & 5 (10\%) & 0 & n/d \\
\emph{BMV/IPPV} & 16 (31\%) & 15 (33\%) & 0.727 \\
\emph{Intubation} & 11 (22\%) & 11 (24\%) & 0.692 \\
\emph{CPR} & 6 (12\%) & 11 (24\%) & 0.315 \\
\emph{CPR and Adrenaline} & 8 (16\%) & 6 (13\%) & 1.0 \\
\emph{CPAP/PEEP} & 4 (8\%) & 1 (2\%) & 1.0 \\
\hline
\end{tabular}
}
\caption{Clinical Characteristics of the newborns in the Public and Validation Datasets. The data is represented as median (interquartile range) or number (\%). The p-value is calculated using the Mann-Whitney U test for quantitative variables and Fisher's Exact test for categorical ones. P-value $<$ 0.05 is considered significant.\newline\textdagger Public dataset: 41 newborns, Validation dataset: 46 newborns.\newline\textdaggerdbl Public dataset: 43 newborns, Validation dataset: 44 newborns.}
\label{tab:clinical_chars}
\end{table}

\subsection{ML Competition}
\label{subsec2.2}
ML competitions require a competition host and several participants. The host provides the competition data and description, the competition’s goals, and potential prizes for the winners. Participants are individuals or groups who build ML models and submit an entry collectively. In our platform, once the host has launched a competition, participants get access to a labelled dataset for training their model (training dataset) and an unlabelled dataset for testing their models (testing dataset). Submitting the labelled predictions constitutes the participant’s entry (submission) into the competition. Participants are allowed multiple submissions throughout the competition with a maximum of 5 submissions per day per participant (or team). While participants have access to both the training and testing datasets, the labels of the testing dataset are automatically hidden from the participants by the platform.  Participants are required to upload their submissions as a single CSV (comma-separated values) file containing the predicted probability and grade (or label) of the developed model. If a validation stage is required, at the end of the competition submissions period, participants must provide an accessible link to their code to assist the competition host in the manual validation process.

\subsection{Validation and Scoring}
\label{subsec2.3}
\begin{figure}[t]
\centering
\begin{minipage}{0.5\textwidth}
\includegraphics[width=\linewidth]{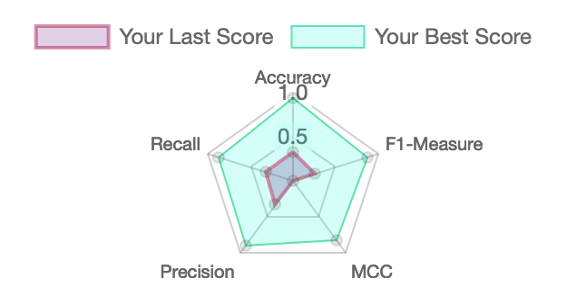}
\end{minipage}%
\begin{minipage}{0.5\textwidth}
\caption{On the public leaderboard, all users can see a visual representation of the scores obtained by each participant for every evaluation measure chosen by the competition host. These representations show the scores relative to the last and best submission of the user.}
\label{fig:side-scoring}
\end{minipage}
\end{figure}

The platform uses a leaderboard system to rank each participant based on their best performing submission in terms of the evaluation measures chosen by the competition host (Figure \ref{fig:side-scoring})

To rank each submission on the public leaderboard, the platform allows for a single performance metric or a weighted sum of different metrics. These metrics vary depending on whether the competition presents a classification or regression (prediction) problem and weights and metrics are selected by the host.

\begin{figure}[t]
\centering
\includegraphics[width=\textwidth]{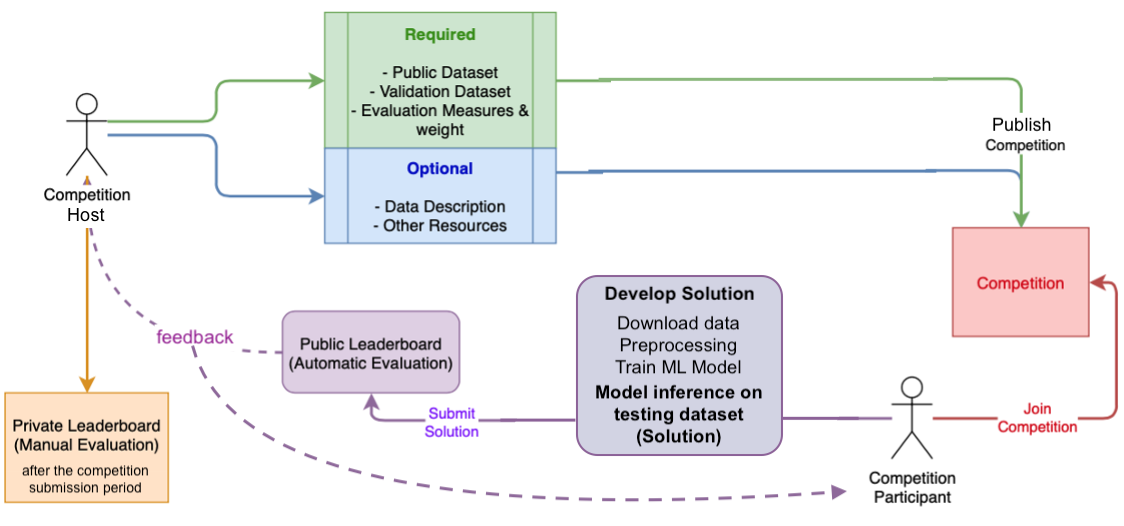}
\caption{Workflow for creating and participting in a ML competition.}
\label{fig:platform-compworkflow}
\end{figure}

In our competition for classifying EEG background patterns, participants were provided with several metrics, including accuracy, F-1 score, a weighted Matthews correlation coefficient (MCC), precision, and recall (Figure: \ref{fig:side-scoring}). Although all performance metrics were accessible to participants, ranking in the leaderboard  was determined using a weighted MCC. This metric was also used to rank the validation performance and the choice of this metric  was kept hidden from the participants to prevent overfitting to a specific metric (Figure: \ref{fig:platform-compworkflow}).

In our competition, the dataset classes were imbalanced. To account for this, we used Matthews correlation coefficient (MCC) \cite{Chicco2020-jn} \cite{Jurman2012-so} \cite{Boughorbel2017-kz}. Additionally, to account for the ordinal nature of the grades, a linear weighting confusion matrix (wCM) was applied to penalises for classification errors beyond 1 grade, is defined as:

\begin{center}
\begin{equation*}
wCM = CM \times
\begin{bmatrix}
1 & 1 & 2 & 3 \\
1 & 1 & 1 & 2 \\
2 & 1 & 1 & 1 \\
3 & 2 & 1 & 1
\end{bmatrix}
\end{equation*}
\end{center}

The Python implementation of the wCM is available with an open-source license on GitHub: \url{https://github.com/fabiom91/MCC\_Weighted/tree/main}. All performance metrics were implemented in python (3.10.4-bullseye) using the scikit-learn v.1.5.1. Confidence Interval (C.I.) 95\%, in parenthesis, calculated using bootstrapping. The participants with the 1st-ranked entry in the private leaderboard were awarded a cash prize of 2000\euro\space from the competition sponsoring (CergenX Ltd, cergenx.com).

\subsection{Submissions}
\label{subsec2.4}
At the conclusion of our competition, we had a total of seven participants, each submitting an average of 27 entries. We selected the top 4 submissions from different participants for manual evaluation on the held-out validation dataset. Each submission used a different ML approach:

\begin{itemize}
    \item Convolutional Neural Network (CNN): based on a previously developed  CNN model from the same author  \cite{Moghadam2022-aw} \cite{Montazeri_Moghadam2022-vt} and fine tuned on the competition data.
    \item ConvNeXt: an off-the-shelf ConvNeXt model, designed for computer vision applications, was used with Gramian angular field transform to encode the time-domain 1D EEG into a 2D image \cite{liu2022convnet} \cite{liu2021swin}.
    \item Extreme Gradient Boosting (XGBoost): quantitative EEG features, extracted using the Matlab NEURAL package \cite{toole2017neural}, were combined with an XGBoost ML model.
    \item Support Vector Machines (SVM): quantitative EEG features extracted using Singular Value Decomposition (SVD), Frank’s copula \cite{10.5555/2811305} and a subset of features from  NEURAL \cite{toole2017neural}, combined using an SVM.
\end{itemize}

The following presents each method in more detail.

\subsubsection{Convolutional Neural Network - CNN}
\label{subsubsec2.4.1}
This submission used a pre-trained CNN-based classifier for grading the EEG that was developed on different EEG dataset \cite{Moghadam2022-aw} \cite{Montazeri_Moghadam2022-vt}. Participants were not excluded from entering the competition if they had already developed algorithms or models in this field.

The method included the following preprocessing. A subset of the available channels were used in a bipolar montage (F3-C3, F4-C4, T3-O1/P3, T4-O2/P4).  Each EEG signal was band-pass filtered between 0.5-32 Hz using a Chebyshev Type 2 bandpass filter, along with a 50 Hz notch filter. Signals were then downsampled to 64 Hz with an anti-aliasing filter and segmented into 1-minute with a 30 second overlap. The signal amplitude was clipped to +/- 250 \(\mu\)V and scaled. The CNN classifier processes a single-channel 1-minute EEG segment and, after 11 layers of processing, provides a vector of probabilities for the four classes as output. The CNN classifier was first trained on a dataset from Helsinki University Hospital, using a different scoring system and a different channel montage to classify EEG background patterns \cite{Moghadam2022-aw}. It was then fine-tuned on the competition dataset using a k-fold (k=3) cross-validation approach with an Adam solver with a learning rate of 1e-5. Within each k-fold, 10\% of the training data were set aside and used to validate the model performance. The fine-tuning process continued for 500 epochs or until the validation loss stopped decreasing for at least 35 epochs. The implementation was done in Python using Keras with a TensorFlow backend and trained on a Geforce GTX 1070 GPU.

Post-processing was used to aggregate the outputs of the 3 models generated from the 3-fold cross-validation, smoothing out spatial and temporal noise in the classifications. Each model generated an output class probability for each processed 1-minute EEG segment from each channel. For spatial smoothing, the output probabilities across channels were combined using an averaging function, and the grade associated with the maximum  probability  determining the predicted grade for each 1-minute epoch. A mild temporal smoothing (moving median filter, window length of 3.5 minutes) was applied to the aggregated CNN output to reduce incidental noise in the time series. Majority voting determined the class for each overlapping set of 120 outputs (1 hour). The final output was achieved through majority voting across the 3 cross-validation models.

\subsubsection{ConvNeXt}
\label{subsubsec2.4.2}
In this submission, the participants treated the EEG classification problem as an image classification task by using a modern, pre-built convolutional network known as ConvNeXt \cite{liu2022convnet} \cite{liu2021swin}], transforming the 1D time-series EEG data into 2D images using Gramian Angular Summation Fields (GASFs) \cite{JMLR:v21:19-763}. The GASFs represent the temporal correlation between each pair of values in the time series. Importantly, GAFs were specifically designed to allow for the use of 1D  time-series data in image classification CNNs, making them an attractive option to use with a ConvNeXt image model. Specifically, the participants used the ConvNeXt-224(tiny) \cite{liu2022convnet2020s} model in their submission.

The EEG data was first filtered with a 0.5 to 12.8 Hz bandpass filter and a 50 Hz notch filter, and then downsampled to 32 Hz. The Root Mean Square (RMS) of the EEG was calculated to use the signal’s average power as the primary feature for classification instead of raw EEG data. This RMS EEG was segmented into windows of length 384 samples and then converted into Gramian Angular Summation Fields (GASFs), resulting in 384 x 384 x 3 images. These images, created from the bipolar montage channels F4-C4, F3-C3, and C4-T4, were resized to 224 x 224 x 3 to fit the ConvNeXt model, which requires 3-channel input (Red, Green, and Blue).

To train this model, all the pre-trained weights of the ConvNeXt-224(tiny) were dropped and the model was trained from scratch on the GASF images of the EEG training dataset.

\subsubsection{Gradient Boosting with NEURAL}
\label{subsubsec2.4.3}
Preprocessing of the EEG included a bandpass filter of 0.5 – 30 Hz to remove the low and high frequency artefacts, followed by downsampling to 64 Hz. The Neonatal EEG feature set in Matlab (NEURAL) package was used to extract the quantitative features from the EEG signals \cite{toole2017neural}. with the following bipolar montage: F3-C4, F3-C3, C4-T4, C3-T3, C4-Cz, Cz-C3, C4-O2/P4 and C3-O1/P3. The EEG signals were then decomposed into four different bandwidth frequencies: [0.5–4; 4–7; 7–13; 13-30] Hz.

The extracted features included four main categories: amplitude, frequency, connectivity, and inter-burst interval features (Table \ref{tab:neural}). Amplitude, frequency, and connectivity features were assessed over short epochs;  these features were computed within a short-time window (64 seconds) that was shifted over time with a 50\% overlap. The median value was used to summarise these features across all epochs and across channels \cite{OTOOLE201742}. Some features were extracted from each frequency band, and some were estimated for the signal before being decomposed (indicated by * in Table 3). In total, 102 features were extracted from each 1-hour epoch.

\begin{table}[htbp]
\centering
\makebox[\textwidth][c]{%
\begin{tabular}{@{}p{4cm} p{4cm} p{4cm} p{4cm}@{}}
\hline
\textbf{Amplitude} & \textbf{Frequency} & \textbf{Connectivity} & \textbf{Inter-burst interval} \\ \hline
\parbox[t]{4cm}{%
\raggedright
$\bullet$ Amplitude total power\\[3pt]
$\bullet$ Amplitude SD\\[3pt]
$\bullet$ Amplitude kurtosis\\[3pt]
$\bullet$ Amplitude skewness\\[3pt]
$\bullet$ Envelope mean\\[3pt]
$\bullet$ Envelope SD\\[3pt]
$\bullet$ rEEG mean\\[3pt]
$\bullet$ rEEG median\\[3pt]
$\bullet$ rEEG lower margin (5th percentile)\\[3pt]
$\bullet$ rEEG upper margin (95th percentile)\\[3pt]
$\bullet$ rEEG width (upper margin -- lower margin)\\[3pt]
$\bullet$ rEEG SD\\[3pt]
$\bullet$ rEEG coefficient of variation\\[3pt]
$\bullet$ rEEG asymmetry (measure of skew about median)%
} &
\parbox[t]{4cm}{%
\raggedright
$\bullet$ Spectral power\\[3pt]
$\bullet$ Spectral relative power (normalized to total spectral power)\\[3pt]
$\bullet$ Spectral flatness\\[3pt]
$\bullet$ Spectral entropy\\[3pt]
$\bullet$ Spectral difference: difference between consecutive spectral estimates\\[3pt]
$\bullet$ Spectral edge frequency: 95\% of spectral power contained between 0.5 and cut-off frequency*\\[3pt]
$\bullet$ Fractal dimension*%
} &
\parbox[t]{4cm}{%
\raggedright
$\bullet$ Brain symmetry index\\[3pt]
$\bullet$ Correlation between envelopes of hemisphere-paired channels\\[3pt]
$\bullet$ Coherence mean value\\[3pt]
$\bullet$ Coherence maximum value\\[3pt]
$\bullet$ Coherence frequency of maximum value%
} &
\parbox[t]{4cm}{%
\raggedright
$\bullet$ 95\% inter-burst interval*\\[3pt]
$\bullet$ 95\% inter-burst interval*\\[3pt]
$\bullet$ Burst percentage*\\[3pt]
$\bullet$ Number of bursts*%
} \\[12pt]
\\ \hline
\end{tabular}%
}
\caption{Overview of EEG features used for seizure prediction, categorized by amplitude, frequency, connectivity, and inter-burst interval measures. Features marked with an asterisk (*) are computed before signal decomposition.}
\label{tab:neural}
\end{table}

A sequential ensemble of decision trees known as extreme gradient boosting machine (XGBoost) was used to develop a predictive model to grade the EEG.  The regularisation parameters were optimised to improve the model’s settings and enhance performance. This process involved systematically exploring various combinations of hyperparameters and selecting the most effective configuration for our specific task. The hyperparameters that were explored included learning rate (0.001, 0.005, 0.01, 0.05, 0.1, 0.2, 0.3), maximum tree depth (2, 3, 4, 5, 6, 7), column sample by tree (0.5, 0.6, 0.7, 0.8, 0.9, 1), gamma values (0, 1, 2), minimum child weight (1, 2, 3), and total number of trees (10, 50, 100, 200, 300, 400, 500). A nested leave-one-out cross-validation (LOOCV) technique was used for parameter optimisation, a reliable procedure for ensuring a comprehensive assessment of the model’s performance without overfitting. Inner LOOCV was used to select the best hyperparameters and an outer LOOCV was employed to evaluate the overall model performance. The  best performing hyperparameters were: max depth = 7, minimum child weight = 1, gamma = 0, learning rate = 0.01, column sample by tree = 0.9 and total number of trees = 500.

\subsubsection{Support Vector Machines (SVM) with quantitative EEG features (qEEG)}
\label{subsubsec2.4.4}
\begin{figure}[t]
\centering
\includegraphics[width=\textwidth]{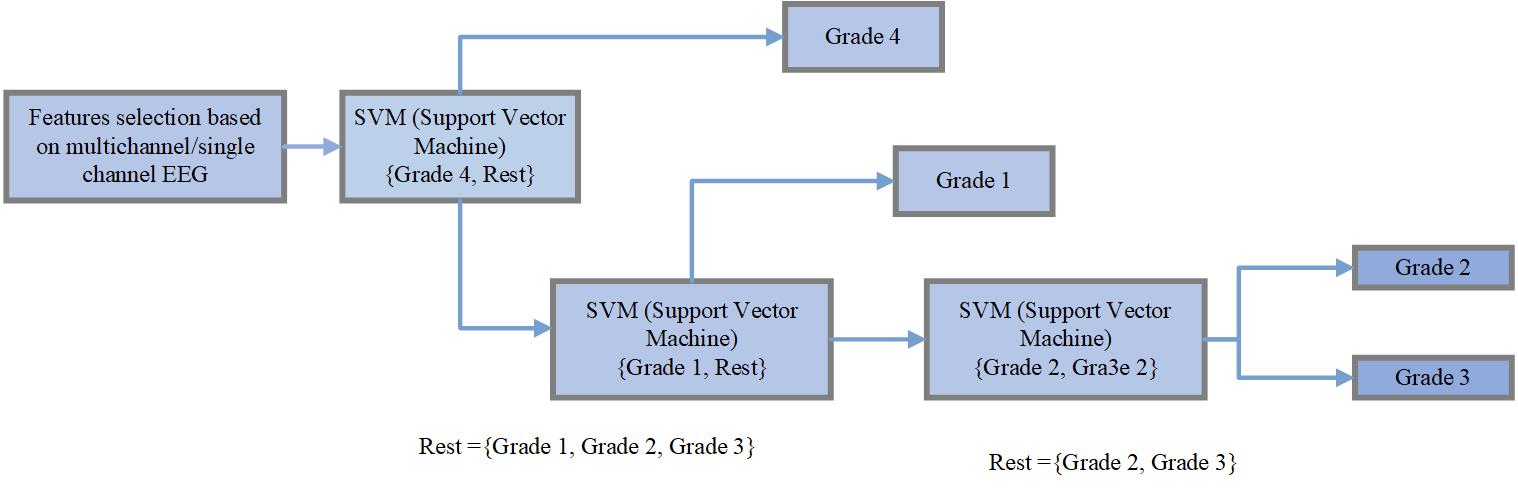}
\caption{Flowchart of the classification of the 4 EEG grades using  an SVM.}
\label{fig:svm}
\end{figure}

The proposed grading system is built using three SVM classifiers. Each classifier is trained using a one-against-rest method, meaning that each classifier learns to identify one specific grade by treating all examples from that grade as positive and all examples from the other grades as negative. The model relies on seven features that are extracted from either multichannel or single EEG channels. (Figure: \ref{fig:svm}).

Singular value decomposition (SVD) was applied across multiple EEG channels to extract features, proving to be an effective method for detecting epileptic seizures in paediatric EEG signals \cite{10283699} \cite{6548330}.The underlying idea is that the maximum singular value in each non-overlapping window would be influenced by the presence of background abnormal patterns. In multichannel EEG data, where each channel consists of non-overlapping segments, different short-duration patterns were identified throughout the recordings \cite{1585} \cite{SKORIC2022108486}.

Since singular values reflect the energy distribution of the data, abrupt changes will impact these values \cite{6548330}. The chosen feature was the maximum singular value across all non-overlapping windows.

Additionally, a copula distribution was used to measure connectivity between EEG channels. Although copulas are commonly used in economics and finance \cite{WEN20121435} \cite{AAS2009182} \cite{doi:10.1198/jasa.2010.tm09572}, their application in biomedical studies, particularly EEG analysis, is relatively rare. Previous studies have suggested the potential of using copulas to analyse EEG time series \cite{ZHENG2022103803} \cite{5495664}.

The copula density function helps visualize signal dependencies, and the copula parameter \(\theta\) quantifies both linear and non-linear relationships between multiple variables \cite{Roy2019} \cite{2006} \cite{Sklar1959FonctionsDR}. For this submission, Frank’s copulas were used to estimate copula parameters \(\theta\), as explicit copulas defined by the closed-form expressions without the distribution constraints \cite{10.5555/2811305}:

\begin{equation*}
C^{(F)}(u_1, u_2) = -\theta^{-1} \cdot \log \left[ 1 + \frac{ (e^{-\theta \cdot u_1} - 1)(e^{-\theta \cdot u_2} - 1) }{ e^{-\theta} - 1 } \right]
\end{equation*}

\begin{itemize}
    \item \(\theta\in(\infty^-,\infty^+ )\), where the parameter \(\theta\) corresponds to the dependency level of the source signals.
    \item \(u_1,u_2\) are the probability integral transform of two channels EEG (e.g. F3, F4).
\end{itemize}

The variables \(u_1\) and \(u_2\) represent short segments (epochs) of EEG data from two channels that have been scaled into a range from \(\theta\) to 1 using a technique called Probability Integral Transform (PIT). By transforming \(u_1\) and \(u_2\) into a uniform scale, the copula can focus on the dependency structure between the two channels without being influenced by their individual characteristics (e.g., amplitude or noise differences).The formula calculates the joint probability (or relationship strength) between \(u_1\) and \(u_2\), which reflects how closely the signals from the two channels are connected.

Other quantitative EEG (qEEG) features included spectral analysis. Spectral power, spectral entropy, and power spectrum were extracted across specific frequency bands according to recommendations for infants aged 32 weeks and above \cite{OToole2016}. Power spectral density (PSD) was estimated using a periodogram with a 50\% overlapping Hamming window. The maximum power value per frequency band was selected as the spectral power feature.

Spectral entropy, which measures the irregularity of a time series in the frequency domain, was calculated using Shannon entropy on the normalized PSD \cite{toole2017neural}. Two additional features were extracted from the range EEG (rEEG): the minimum value of the 95th percentile of rEEG within each frequency band and the maximum range of rEEG, defined as the difference between the 95th and 5th percentiles within each band.

MATLAB’s implementation of a sequential minimal optimization (SMO) based algorithm for training linear SVMs was used.

\section{Results}
\label{sec3}
All performance metrics were implemented in python (3.10.4-bullseye) using the scikit-learn v.1.5.1. Confidence Interval (C.I.) 95\%, in parenthesis, calculated using bootstrapping.

% PUBLIC LEADERBOARD TABLE
\begin{table}[t]
\makebox[\textwidth]{
\begin{tabular}{>{\raggedright}m{0.2\textwidth} 
                *{5}{>{\centering\arraybackslash}m{0.2\textwidth}}}
\hline
\textbf{Method} & \textbf{Weighted MCC} & \textbf{Accuracy} & \textbf{F1} & \textbf{Precision} & \textbf{Recall} \\
\hline
\textbf{XGBoost\\NEURAL-toolbox} & 
\begin{tabular}[c]{@{}c@{}}0.761\\(0.618--0.909)\end{tabular} & 
\begin{tabular}[c]{@{}c@{}}0.891\\(0.812--0.953)\end{tabular} & 
\begin{tabular}[c]{@{}c@{}}0.843\\(0.710--0.934)\end{tabular} & 
\begin{tabular}[c]{@{}c@{}}0.873\\(0.750--0.963)\end{tabular} & 
\begin{tabular}[c]{@{}c@{}}0.825\\(0.708--0.937)\end{tabular} \\

CNN & 
\begin{tabular}[c]{@{}c@{}}0.713\\(0.566--0.864)\end{tabular} & 
\begin{tabular}[c]{@{}c@{}}0.859\\(0.781--0.937)\end{tabular} & 
\begin{tabular}[c]{@{}c@{}}0.768\\(0.607--0.900)\end{tabular} & 
\begin{tabular}[c]{@{}c@{}}0.767\\(0.621--0.936)\end{tabular} & 
\begin{tabular}[c]{@{}c@{}}0.813\\(0.693--0.915)\end{tabular} \\

ConvNeXt & 
\begin{tabular}[c]{@{}c@{}}0.677\\(0.522--0.835)\end{tabular} & 
\begin{tabular}[c]{@{}c@{}}0.859\\(0.781--0.937)\end{tabular} & 
\begin{tabular}[c]{@{}c@{}}0.807\\(0.677--0.901)\end{tabular} & 
\begin{tabular}[c]{@{}c@{}}0.862\\(0.726--0.965)\end{tabular} & 
\begin{tabular}[c]{@{}c@{}}0.775\\(0.662--0.881)\end{tabular} \\

SVM\\qEEG features & 
\begin{tabular}[c]{@{}c@{}}0.437\\(0.257--0.627)\end{tabular} & 
\begin{tabular}[c]{@{}c@{}}0.750\\(0.641--0.844)\end{tabular} & 
\begin{tabular}[c]{@{}c@{}}0.568\\(0.328--0.729)\end{tabular} & 
\begin{tabular}[c]{@{}c@{}}0.821\\(0.292--0.879)\end{tabular} & 
\begin{tabular}[c]{@{}c@{}}0.557\\(0.372--0.732)\end{tabular} \\
\hline
\end{tabular}
}
\caption{Public Leaderboard, automatically generated by the competition platform comparing the true labels of the testing dataset against the one submitted by the participants.}
\label{tab:public_leaderboard}
\end{table}

% PRIVATE LEADERBOARD TABLE
\begin{table}[t]
\makebox[\textwidth]{
\begin{tabular}{>{\raggedright}m{0.2\textwidth} 
                *{5}{>{\centering\arraybackslash}m{0.2\textwidth}}}
\hline
\textbf{Method} & \textbf{Weighted MCC} & \textbf{Accuracy} & \textbf{F1} & \textbf{Precision} & \textbf{Recall} \\
\hline
\textbf{CNN} & 
\begin{tabular}[c]{@{}c@{}}0.378\\(0.273--0.481)\end{tabular} & 
\begin{tabular}[c]{@{}c@{}}0.576\\(0.5--0.647)\end{tabular} & 
\begin{tabular}[c]{@{}c@{}}0.615\\(0.539--0.680)\end{tabular} & 
\begin{tabular}[c]{@{}c@{}}0.689\\(0.617--0.755)\end{tabular} & 
\begin{tabular}[c]{@{}c@{}}0.613\\(0.533--0.689)\end{tabular} \\

ConvNeXt & 
\begin{tabular}[c]{@{}c@{}}0.350\\(0.260--0.444)\end{tabular} & 
\begin{tabular}[c]{@{}c@{}}0.571\\(0.5--0.641)\end{tabular} & 
\begin{tabular}[c]{@{}c@{}}0.551\\(0.453--0.637)\end{tabular} & 
\begin{tabular}[c]{@{}c@{}}0.634\\(0.531--0.722)\end{tabular} & 
\begin{tabular}[c]{@{}c@{}}0.534\\(0.450--0.614)\end{tabular} \\

XGBoost\\NEURAL-toolbox & 
\begin{tabular}[c]{@{}c@{}}0.311\\(0.219--0.404)\end{tabular} & 
\begin{tabular}[c]{@{}c@{}}0.489\\(0.413--0.560)\end{tabular} & 
\begin{tabular}[c]{@{}c@{}}0.491\\(0.407--0.560)\end{tabular} & 
\begin{tabular}[c]{@{}c@{}}0.524\\(0.448--0.606)\end{tabular} & 
\begin{tabular}[c]{@{}c@{}}0.566\\(0.483--0.647)\end{tabular} \\

SVM\\qEEg features &
\begin{tabular}[c]{@{}c@{}}0.240\\(0.160--0.325)\end{tabular} & 
\begin{tabular}[c]{@{}c@{}}0.402\\(0.331--0.473)\end{tabular} & 
\begin{tabular}[c]{@{}c@{}}0.420\\(0.323--0.496)\end{tabular} & 
\begin{tabular}[c]{@{}c@{}}0.526\\(0.409--0.635)\end{tabular} & 
\begin{tabular}[c]{@{}c@{}}0.481\\(0.404--0.552)\end{tabular} \\
\hline
\end{tabular}
}
\caption{Private Leaderboard, manually generated after running each submission model inference on the hidden validation dataset and comparing true and predicted labels.}
\label{tab:private_leaderboard}
\end{table}

On the public leaderboard in Table \ref{tab:public_leaderboard}, the XGBoost model performed best with 0.76 weighted MCC, followed by the CNN (0.71), ConvNeXt (0.68) and SVM (0.43) methods. However, in comparison, all methods had relatively poor performance on the validation dataset—see Table \ref{tab:private_leaderboard}. Ranking of methods also changed from the XGBoost to the CNN as the best model across all metrics. Next best was the imaged-based ConvNeXt model with 0.35 weighted MCC, then followed by the XGBoost (0.31) and SVM (0.24) methods

\begin{figure}[t]
\centering
\includegraphics[width=0.24\textwidth]{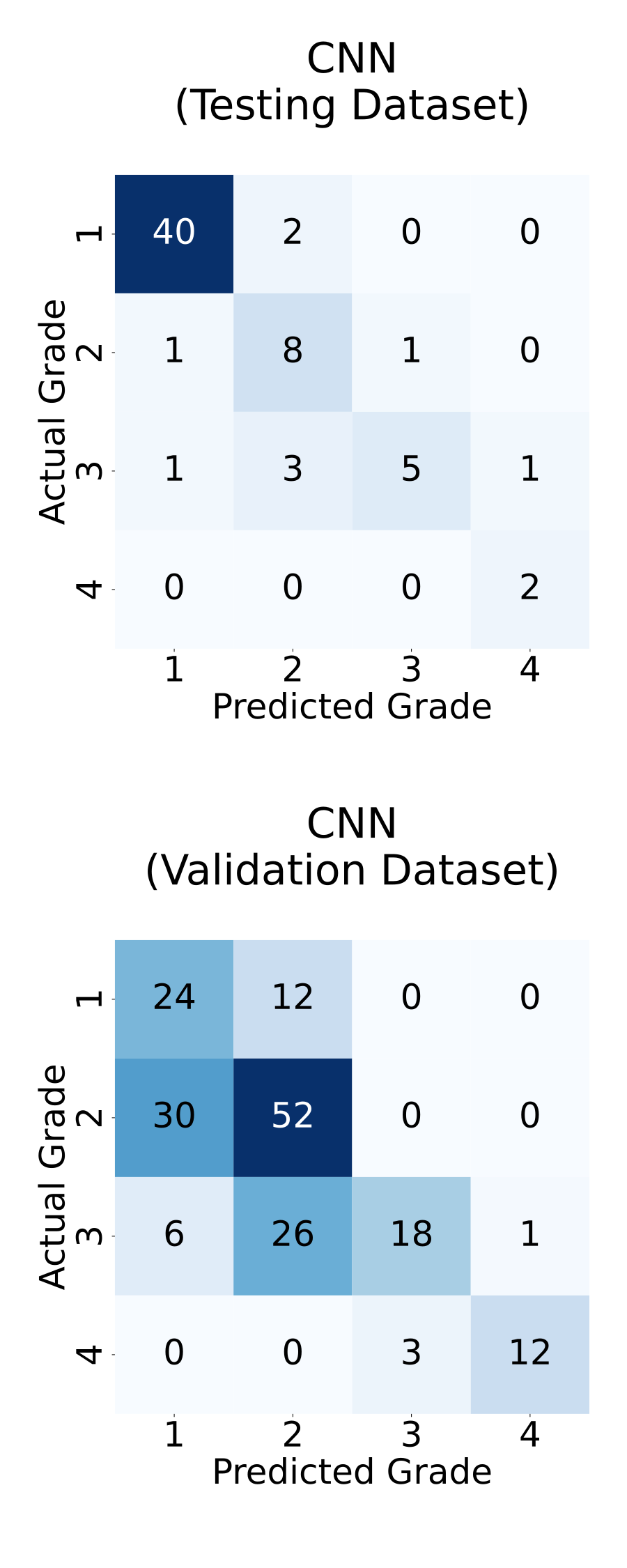}%
\hfill
\includegraphics[width=0.24\textwidth]{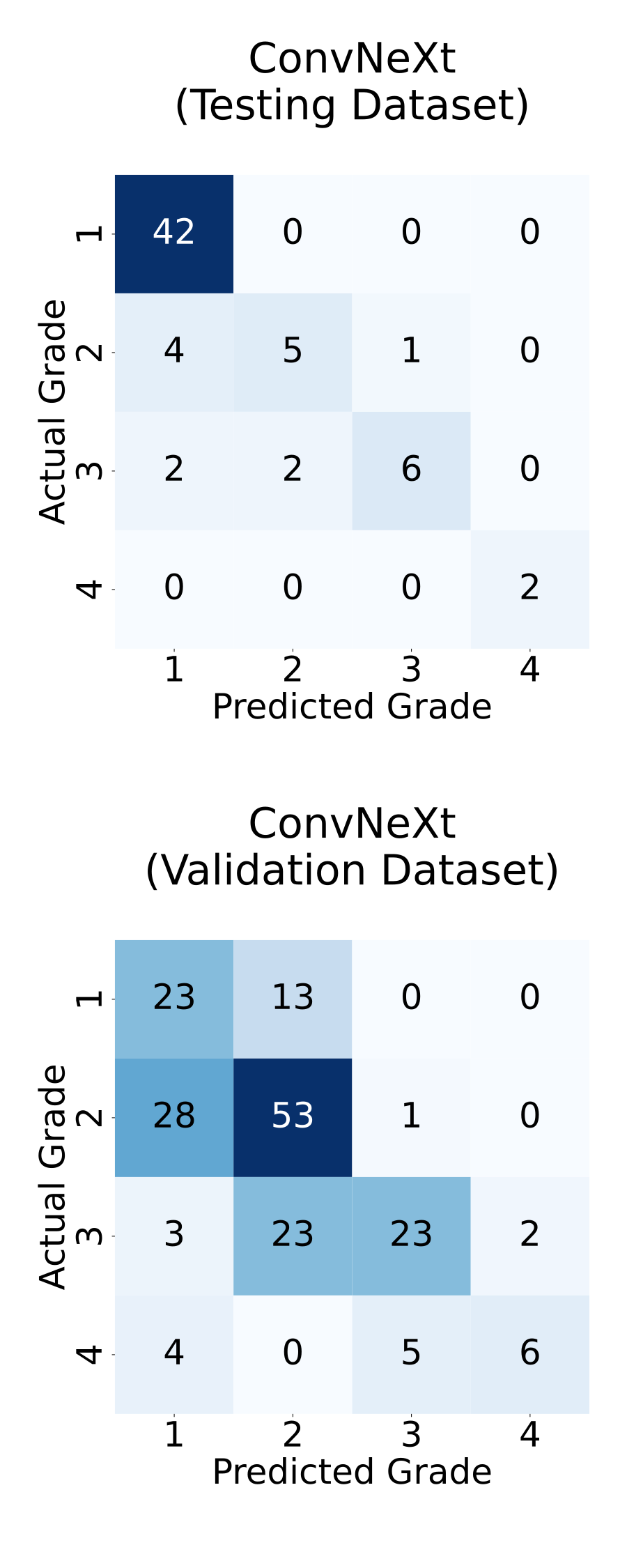}%
\hfill
\includegraphics[width=0.24\textwidth]{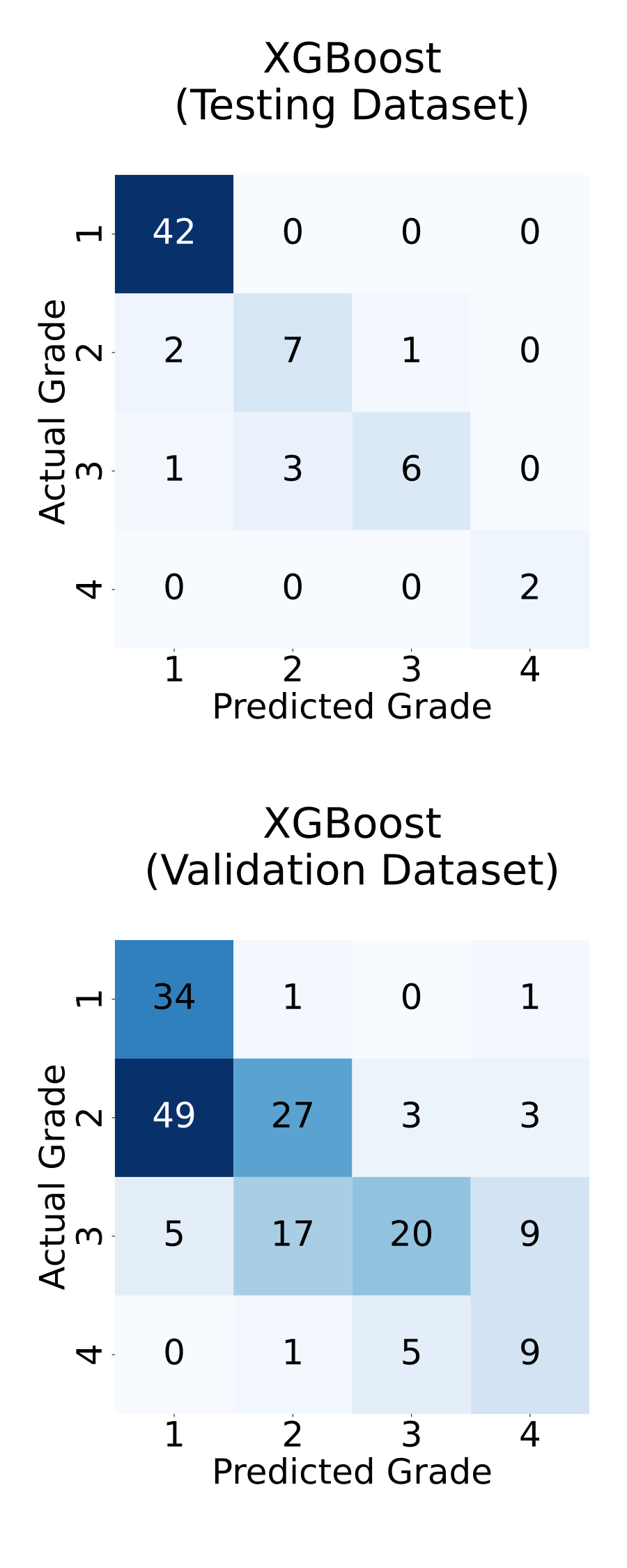}%
\hfill
\includegraphics[width=0.24\textwidth]{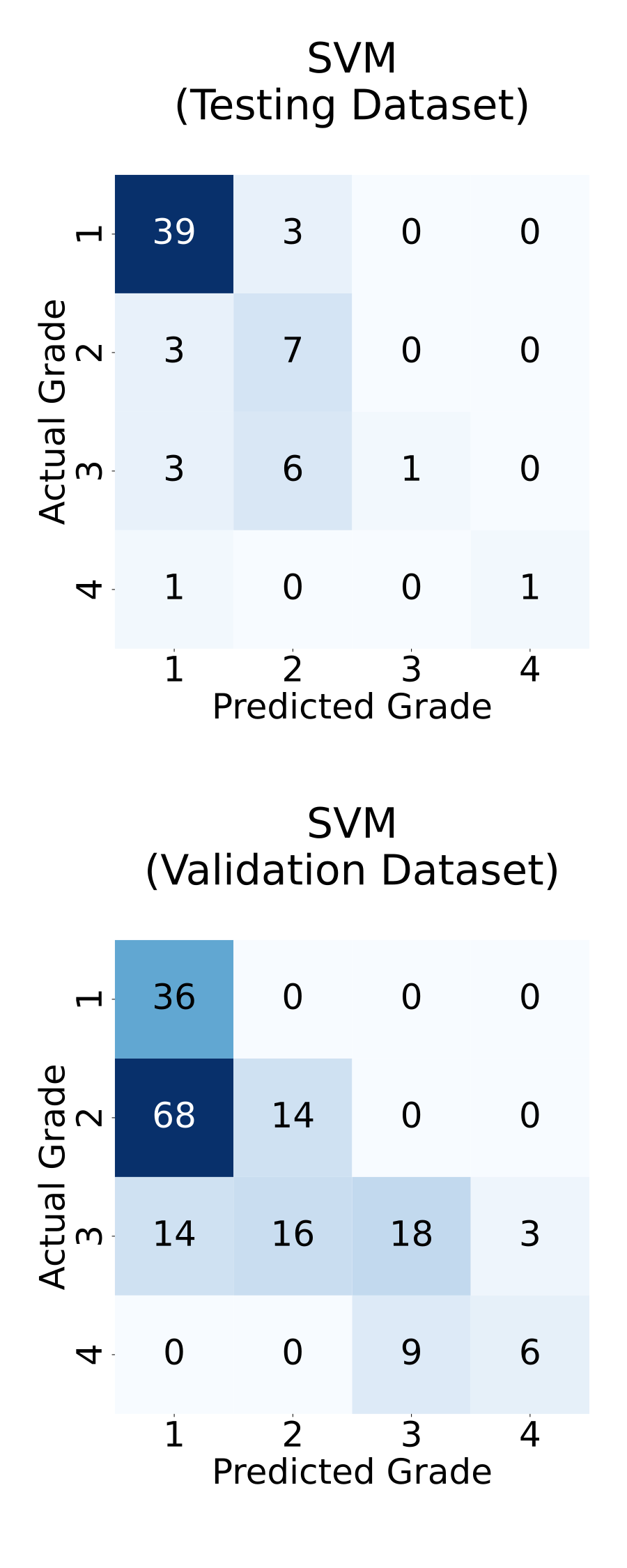}
\caption{Confusion matrices for each model (CNN, ConvNeXt, XGBoost, and SVM) on both the testing and validation datasets. The numbers within the matrices represent the number of epochs. All models exhibited higher accuracy in predicting grade 1 in the testing datasets. Misclassification between grades 1 and 2 was pervasive across models, particularly in the validation dataset, as discriminating between these grades can prove challenging, even for expert neurophysiologists.}
\label{fig:confusion_matrices}
\end{figure}

Upon reviewing the overall trend of misclassifications across all models, we can see a tendency for grades to be underestimated. This is consistently observed in the confusion matrices, where a higher concentration of values appears in the area below the diagonal in each matrix.

The SVM model encountered the most difficulty distinguishing between grades 1 and 2, followed by XGBoost, indicating a superior discriminative ability for deep learning models. In the validation set, grade 3 obtained the highest incidence of incorrect classifications, with misclassifications ranging from grade 1 to 4. Notably, grade 4 classifications displayed improved performance over grade 3, as these particularly severe (inactive) EEG patterns are comparatively easier to discern  (Figure \ref{fig:confusion_matrices}).

\begin{figure}[t]
\centering
\includegraphics[width=\textwidth]{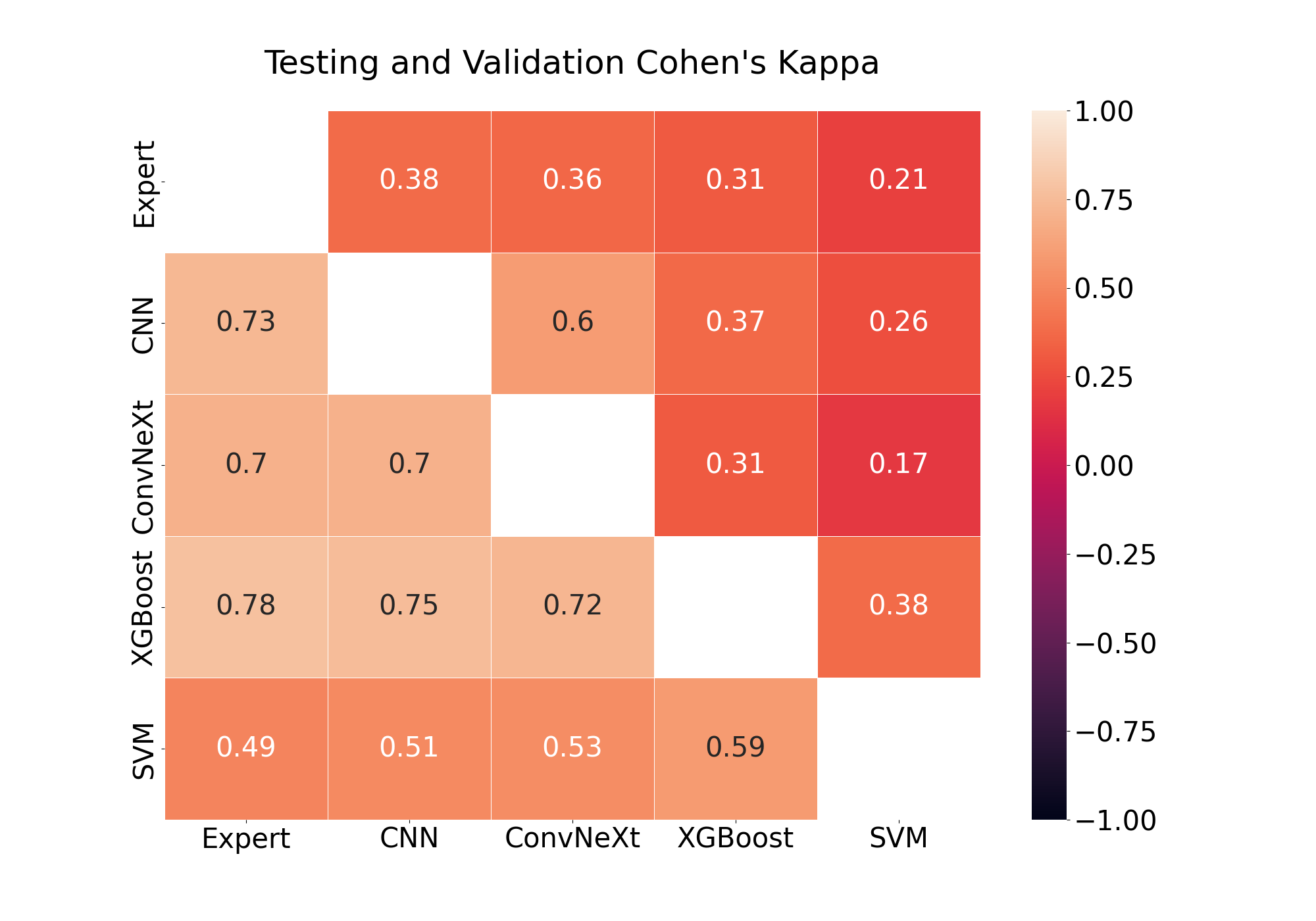}
\caption{Agreement among the different methods as measured by Cohen’s Kappa and agreement with the human experts annotations on the hidden validation dataset (upper half above the diagonal) and the testing dataset (lower half below the diagonal).}
\label{fig:cohenK}
\end{figure}

\begin{table}[t]
\makebox[\textwidth]{
\begin{tabular}{>{\raggedright}m{0.14\textwidth} 
                >{\centering}m{0.14\textwidth}
                *{5}{>{\centering\arraybackslash}m{0.16\textwidth}}}
\hline
 & \textbf{Method} & \textbf{Weighted MCC} & \textbf{Accuracy} & \textbf{F1-Measure} & \textbf{Precision} & \textbf{Recall} \\
\hline
Testing Dataset & 
\multirow{2}{*}{\makecell{Ensemble\\(majority\\vote)}} & 
\begin{tabular}[c]{@{}c@{}}0.695\\(0.533--0.861)\end{tabular} & 
\begin{tabular}[c]{@{}c@{}}0.875\\(0.797--0.938)\end{tabular} & 
\begin{tabular}[c]{@{}c@{}}0.834\\(0.705--0.933)\end{tabular} & 
\begin{tabular}[c]{@{}c@{}}0.913\\(0.825--0.979)\end{tabular} & 
\begin{tabular}[c]{@{}c@{}}0.800\\(0.686--0.903)\end{tabular} \\[12pt]

Validation Dataset & & 
\begin{tabular}[c]{@{}c@{}}0.366\\(0.269--0.461)\end{tabular} & 
\begin{tabular}[c]{@{}c@{}}0.516\\(0.440--0.592)\end{tabular} & 
\begin{tabular}[c]{@{}c@{}}0.564\\(0.473--0.640)\end{tabular} & 
\begin{tabular}[c]{@{}c@{}}0.651\\(0.568--0.720)\end{tabular} & 
\begin{tabular}[c]{@{}c@{}}0.595\\(0.513--0.669)\end{tabular} \\
\hline
\end{tabular}
}
\caption{Performance metrics of the Ensemble (majority vote) of the 4 submissions in the leaderboard for the testing and validation datasets.}
\label{tab:ensemble_performance}
\end{table}

To test for similarity among the 4 methods, we calculate pairwise Cohen’s kappa for all methods groups (Figure: \ref{fig:cohenK}). We find that agreement among methods is moderate, with the highest pairwise agreement between the CNN and XGBoost methods for testing (k=0.75) whereas the highest agreement between the CNN and ConvNeXt methods in validation sets (k=0.60).  We also generate a new model from the  ensemble of the 4 methods using majority vote. Although the ensemble had performance close to the best performance methods, it failed to yield better performance than the best models for each dataset; see Table \ref{tab:ensemble_performance}.

\section{Discussion}
\label{sec4}
The competition results highlighted the practicality of crowdsourcing ML development and sharing anonymized EEG data. We evaluated various ML algorithms by their ability to classify the severity of abnormalities in EEG background patterns. While XGBoost achieved the top score on the public leaderboard, the CNN and ConvNeXt methods outperformed it on the held-out validation set. This indicates that in our competition, Deep Learning (DL) algorithms tend to generalize better than traditional ML techniques, which relied on features extracted from the EEG signal. The CNN architecture, which unlike all other methods was trained on the raw EEG, performed best on the validation set. However, the method had a notable advantage over other methods due to its development prior to the competition. It was pre-trained on a similar dataset and subsequently fine-tuned using the competition’s training dataset \cite{Moghadam2022-aw} \cite{Montazeri_Moghadam2022-vt} \cite{Kota2024} \cite{Montazeri2024}. The success of the pre-trained model underscores the versatility of pre-trained models in addressing novel tasks with limited domain-specific data.

We observed substantial performance deterioration for every model when comparing the public (testing) and private (validation) leaderboards. This discrepancy is likely due to the small dataset size and large class imbalance in the training and testing sets therefore leading to overfitting. The drop in performance may be associated with the decrease in the distribution of grade 1 in the validation dataset (20\%) compared to the training (59\%) and testing (66\%) datasets. The large class imbalance in the training and testing sets may have biased the models towards underestimating the severity of EEG grade, a hypothesis supported by the biased validation testing in (Figure: \ref{fig:confusion_matrices}). None of the methods accounted for class imbalance during training and therefore this is a likely cause, at least in part, to the poor validation performance.

The gap between testing and validation performance underscores the necessity for high-quality, diverse, and accessible datasets to train ML models to generalise effectively in real-world clinical contexts. Generalisation to unseen data is essential for the clinical application of these algorithms, especially in fast-paced environments like neonatal intensive care units, where infants' physio-pathological conditions and clinical practices may vary from those present during model training. These insights highlight the critical need to tackle prevalent challenges in machine learning research in this field, including limitations in dataset size, class imbalance, and reproducibility.

It must be noted that the scope of this study is not to provide clinical validation of ML models but to test the concept of running ML competitions to crowdsource model development and incentivize data sharing. None of the ML models in this study have been developed to be adopted cot-side by clinicians in real-world scenarios.

The best-performing model in our competition utilized only four EEG channels from an 8-channel bipolar montage, reaffirming findings by Raurale et al. (2021) that reduced-channel systems can maintain robust performance \cite{Raurale2021}. This approach holds promise for improving accessibility in resource-limited settings, where multi-channel EEG systems may not be available. Further validation of reduced-channel configurations is necessary to confirm their clinical utility without sacrificing significant diagnostic accuracy as a reduced channel montage has obvious practical advantages. Additionally, techniques that require fewer EEG channels could be useful for amplitude-integrated EEG (aEEG) systems, the most common form of EEG monitoring available in NICUs, therefore making them available to a broader group of clinicians and research centres.

Although ConvNeXt is a more recent iteration of CNN, the specific ConvNeXt model used relied on a pre-designed framework for images, necessitating the transformation of the 1D EEG time-series signal into image format. This process, which prevents the end-to-end optimisation potential of deep learning, could introduce information loss or artifacts due to the compression of the signal, negatively affecting performance. While ConvNeXt's performance was only slightly lower than that of the top-performing model, which also used a CNN, the winning model was specifically tailored for a similar application of grading Neonatal EEG and it was pre-trained on private neonatal EEG datasets. This suggests that while off-the-shelf deep learning models like ConvNeXt can be effective, application-specific models may offer better results.

Due to the observed drop in performances in the validation dataset across all methods, it is likely that models trained with a dataset of this scale  (50 newborns) are insufficient to generalise to the population of at-risk newborns and therefore not suitable for clinical adoption. Several other considerations must also be addressed when designing a cot-side model, including privacy, maintainability, and computational efficiency. Models capable of running on edge devices offer privacy advantages, as data remains within the device hardware. On the other hand, hardware constraints can limit both inference speed and potential model complexity, and consequently, accuracy. Furthermore, updating these models on edge devices can be challenging, which means these models cannot easily take advantage of updates or refinements based on new data. In contrast, cloud-based models allow for seamless remote updates and avoid the hardware limitations of edge devices. Still, they require strict security protocols to ensure data protection and a reliable internet connection.  Modern deep learning methods, trained on larger datasets, typically demand more advanced hardware than traditional machine learning methods. However, traditional approaches often involve computationally intensive preprocessing for feature extraction before they can make predictions on new data. Therefore, it is important to analyse the costs and benefits of each approach when deciding which machine learning method to use.

The benefits of ML competitions are evident for both hosts and participants. By hosting a competition, hosts gain motivation to share their data, creating a promotional platform for their dataset among experts. This approach can leverage a diverse talent pool for developing ML models. ML competitions provide participants with access to well-defined datasets and enable peer learning through competition and collaboration. Whether working individually or partnering with others in the competition, these events facilitate knowledge sharing and professional cooperation. Furthermore, winners gain from prizes, heightened visibility, and the chance to refine their ML skills. For hosts and participants alike, ML competitions serve as a means to cultivate a community of experts that encourages ongoing professional collaboration and the sharing of data and expertise across various fields.

Future efforts should focus on increasing the amount of available data for model training and validation, refining automated model evaluation techniques, and promoting reproducibility to ensure that machine learning tools can reliably aid in neonatal care. By continuing to harness collective expertise through competitions and collaborative research, we can make significant strides toward improving diagnostic accuracy and therefore patient outcomes in neonatal healthcare.

There were some limiting factors running this study. Some of these limitations were due to the limited data available for training and testing the methods. The public data represented a subset of the data gathered from a multi-centre study. Although this dataset is large for neonatal EEG studies (169 hours from 53 newborns), it only consisted of data from one centre (EEG recordings from Cork University Maternity Hospital).

ML competitions inherently have limitations, mainly due to their competitive nature and the requirement for scoring submissions. A single performance metric must be used to score and rank each submission and ultimately to declare a competition winner. However, these metrics might overlook other important aspects of the model, such as inter-rater agreement, clinical utility, or computational efficiency. Additionally, the score is closely related to the scope of the testing and validation data—for example the distribution of classes, aetiologies of the newborn, and clinical centre where the EEG was recorded—which can impact the ability to thoroughly assess a model’s generalisation performance on unseen data. All models should also be evaluated regarding the feasibility of clinical implementation and in relation to inter-rater agreement between the EEG experts.

Additionally, research indicates variability in inter-rater agreement regarding expert classifications of EEG background patterns in newborns with HIE. Wusthoff (2017) \cite{Wusthoff2017aa} reported the highest Cohen-K coefficient of 0.7 (p $<$ 0.001) using a five-grade scale. In contrast, Shavonne (2019)\cite{Massey2019} emphasised the necessity for educational and technological strategies to enhance performance in EEG background scoring. These findings reveal a degree of discretion in grading EEG background patterns in newborns with HIE, which affects the ability of a model to generalise to unseen data. This issue can be lessened by training models on extensive and varied datasets labelled by multiple experts. However, adopting a grading standard that minimises human subjectivity might be a more effective approach.

Lastly, the necessity of funding or sponsorship needed to prize the challenge competitively with larger platforms like Kaggle and the outreach efforts involved in advertising the competition can be a challenge for research projects when competing with large industry-sponsored projects. Despite our relative resource constraints, the competition successfully received 187 submissions, although it attracted only seven teams during its relatively brief submission period. The competition was promoted through our centre’s ( INFANT Research Centre, University College Cork) internal channels and social media in addition to promotion through the AI-4-NICU COST (European Cooperation in Science and Technology) Action network, but our reach to the broader public was limited without a dedicated advertising team. To encourage public participation, a video campaign was launched on social media channels, including Facebook, Instagram, LinkedIn, X, and YouTube, before the competition’s start, and a 2,000\euro prize was offered to the first-place winner. Future competitions with longer durations and larger prizes could attract more participants providing a more comprehensive benchmark for ML methods in classifying neonatal EEG abnormalities. Overall, this experience serves as a successful pilot study for future endeavours.

\subsection{Data Availability}
\label{subsec4.1}
The data of the competition discussed in this paper is published on Zenodo and available with open-access license \cite{o_toole_2022_6587973}.

In accordance with the General Data Protection Regulation (GDPR (EU) 2016/679), users have the right to access their personal data collected by the AI competition platform operated by INFANT and request its permanent deletion from our servers/database.

\subsection{Software Availability}
\label{subsec4.2}
The Platform’s open-source code is stored in a public GitHub repository: \url{https://github.com/fabiom91/AI\_Competition\_Platform} with a BSD License.

The competition participants have agreed to share their code with an open-source license on their repositories available online:
\begin{itemize}
    \item CNN: \url{https://github.com/smontazeriUH/INFANT-Data-Science-Challenge.git}
    \item ConvNext: \url{https://github.com/fabiom91/ConvNeXt\_AI\_Competition\_Platform}
    \item Gradient Boosting with NEURAL: \url{https://github.com/Minimnim/INFANT\_ML\_challenge}
    \item SVM with qEEG: \url{https://github.com/tamaraceranic/Support-Vector-Machines-SVM-with-quantitative-features-EEG-qEEG-}
\end{itemize}

\section{Author Contributions}
\label{sec5}
John O’Toole and Geraldine B. Boylan led the study's conception and design. Additionally, John O’Toole and Fabio Magarelli managed the data formatting for the ML competition. Fabio Magarelli was responsible for developing and deploying the competition platform, overseeing the competition, and evaluating the algorithms against the validation data. He also wrote the manuscript's initial draft. Geraldine B. Boylan was the principal investigator in the clinical trial that gathered the EEGs used in the competition. Saeed Montazeri, Feargal O’Sullivan, Dominic Lightbody, Minoo Ashoori, and Tamara Skoric Ceranic developed the ML models for the competition. All authors conducted a critical review and revision of this manuscript.

\section{Competing Interests}
\label{sec6}
Geraldine B. Boylan has a consultancy with ReAlta Life Sciences and Nihon Kohden. She is the co-founder of UCC  Spin out company CergenX Ltd and start-up company Kephala Ltd. John O'Toole is Head of AI Research at CergenX Ltd. 

\section{Grant Information}
\label{sec7}
This work was supported by an Innovator Award and a Strategic Translational Award from the Wellcome Trust (209325 and 098983). The article is based upon work by the COST Action AI-4-NICU (CA20124), supported by COST (European Cooperation in Science and Technology, https://www.cost.eu/). This publication has emandated from research conducted with the financial support of Taighde Éireann-Research Ireland under Grant No. 18/CRT/6223.

\section{Acknowledgments}
\label{sec8}
We wish to acknowledge the work of Sean R. Mathieson and Jerry Deasy from INFANT Research Centre. They respectively have taken part in the consensus grading of the EEG data used in the competition and helped set up the remote Ubuntu server for the deployment of the Platform.

\end{document}